\documentclass[%
 reprint,
 amsmath,amssymb,
 aps,
prl,
]{revtex4-2}

\usepackage{graphicx}
\usepackage{dcolumn}
\usepackage{bm}

\begin{document}

\preprint{APS/123-QED}

\title{Nanomechanically Induced Transparency}

\author{E. C. Diniz}
 \affiliation{Universidade do Estado de Mato Grosso, Campus Tangará da Serra, Rodovia MT 358, Km 07 (s/n), Jardim Aeroporto, Tangará da Serra, Mato Grosso, Brazil. }
\author{O. P. de S\'a Neto}%
 \email{olimpiopereira@phb.uespi.br}
\affiliation{Coordena\c{c}\~ao de Ci\^encias da Computa\c{c}\~ao, Universidade Estadual do Piau\'i, Campus Professor Alexandre Alves de Oliveira, 64202--220, Parna\'iba, Piau\'i, Brazil.}%

\date{\today}

\begin{abstract}
In this paper, we investigate a nanomechanically induced transparency (NIT) effects that arises from the coupling of a nanoelectromechanical system and a trapped ion. By confining the ion in mesoscopic traps and capacitively coupling it with a nanoelectromechanical system suspended as electrodes, the research is intricately focussed on the implications of including the ion's degrees of freedom. The Lamb-Dicke approximation is crucial to understanding the effects of phonon exchange with electronic qubits and revealing transparency phenomena in this unique coupling. The results underline the importance of the Lamb-Dicke approximation in modelling the effects of transparency windows in nanoelectromechanical systems. 
\end{abstract}

\maketitle

\section{Introduction}
Contemporary scientific research has witnessed several advances, especially in the field of nanotechnology and electromechanical systems. The Colombian interaction between nanoelectromechanical systems and trapped ions stands out even more for its intrinsic rewriting algebra of Quantum Electrodynamics (QED)\cite{haroche2013nobel,raimond2001manipulating}. This convergence of phenomena offers a unique perspective that could revolutionise technology and open up new frontiers in the understanding of quantum state engineering \cite{dowling2003quantum,Ospelkaus11,Timoney11}.

Nanoelectromechanical systems (NEM) represent an emerging frontier in the miniaturisation of mechanical devices \cite{bowen2015quantum}. With dimensions in the order of nanometres, these systems exhibit unique properties, such as high sensitivity and responsiveness at tiny scales. Trapped ions, on the other hand, are charged particles confined in electromagnetic fields, usually generated by ion traps. These traps offer precise control over the movements and quantum states of the ions, making them ideal components for studying quantum systems and quantum computing applications.  Colombian interaction \cite{jackson2021classical,de2022temperature}, in this context, refers to the electrostatic attraction between a charged particle and a charged plate, i.e. a crucial phenomenon when dealing with trapped ions and a NEM.
In the context of the Columbian interaction between NEM and trapped ions, transparency can be achieved by precisely controlling the oscillation of the NEM, resulting in a controlled manipulation of the mechanical response of the materials involved \cite{souza2015electromagnetically}.

From an experimental point of view, investigating the presence and role of quantum coherence in a wide variety of systems has attracted considerable interest over the past years \cite{Streltsov2017}. In view of these stimulating results, here we examine the quantum coherence response for a less explored experiment: an ion trap coupled to NEM.

The first proposal of such a system was made by Tian and Zoller \cite{tian04}, with ions in mesoscopic traps, while suspended nanomechanical resonators played the role of tiny electrodes of the trap, acting as nanomechanical resonators with their own degrees of freedom.

Additionally, Hensinger \textit{et al.} \cite{hensinger05} showed that a single trapped ion could be used to test the quantum nature of a mesoscopic mechanical oscillator. The investigation of quantum coherence in mechanical interactions has great relevance for quantum information processing, with proof of experimental feasibility 
and generation of entanglement \cite{nicacio2013motional,bentley2014detection,Kielpinski12}.

The article is organised as follows. First, we review the NEM-ion trap coupling, modelling and formulating the interaction Hamiltonian for dynamical calculations of a single-phonon system on command.  We then analyse the open quantum dynamics of the system that effectively describe the NEM-ion trap coupling. Finally, we obtain the spectra and analyse the induced transparency effects, thereby closing our results and presenting our conclusions.

\section{The NEM-Ion Trap Coupling}
\begin{center}
\begin{figure}[h]
\begin{center}
\includegraphics[scale=0.3]{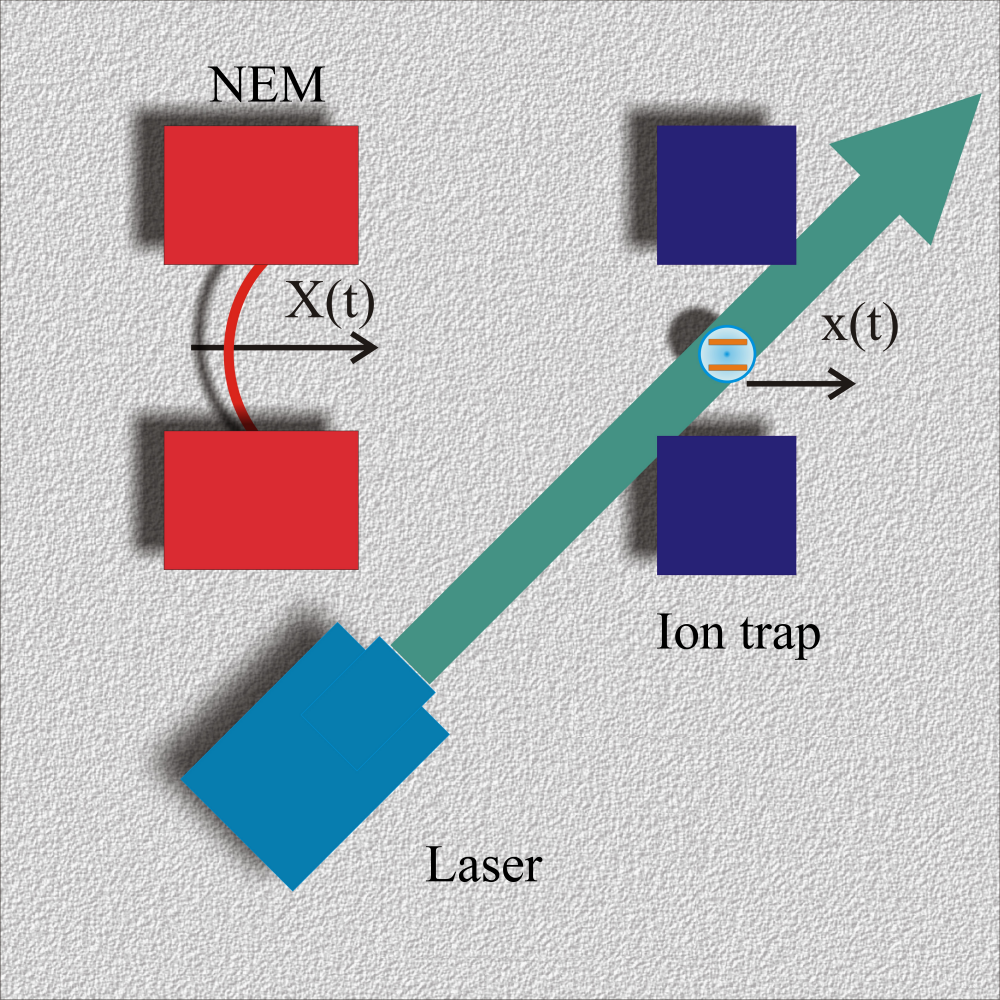}
\caption{Schematic model of experiment: vibrational mode of trapped Ion interacting with the electric field of a laser, and electrostatically with a nanoelectromechanical system.}
\label{nemsion0}
\end{center}
\end{figure}
\end{center}

The interaction of a single ion trap and a NEM is based on the experiment proposed in Ref.\,\cite{hensinger05}; see, e.g., an illustration of this Ion-NEM coupling in Fig.\,\ref{nemsion0}. 

Given this, the corresponding electrostatic energy of such an Ion-NEM system reads
\begin{equation}\label{eq1}
V_{c}= k \frac{V_0C_0}{d+X(t)-x(t)},
\end{equation}
where $C_{0}$ denotes the capacitance of the gate, $V_0$ represents the voltage bias, $d$ corresponds the separation 
of the center of mass of the ion and NEM from equilibrium, and $x(X)$ is the position of the ion 
(NEM). For small oscillations ($X, x<<d$), expanding up to second order, we have \cite{hensinger05}
\begin{eqnarray}\label{eq2}
V_{c}= -\chi X(t)x(t),
\end{eqnarray}
where $\chi= 2keV_{0}C_{0}$ and the first-order term is neglected due to the rapid oscillations.

Now, defining 
$X = \sqrt{ \frac{\hbar}{2M\omega} } \left( a + a^{\dag} \right)$,
and
$x = \sqrt{ \frac{\hbar}{2m\nu} } \left( b + b^{\dag} \right)$,
with $M(m)$ being the mass and $\omega(\nu)$ being the frequency of NEM(Ion) respectively. The Hamiltonian of the whole system can be written as
\begin{equation}\label{eq3}
H=\hbar \omega a^{\dagger} a+ \hbar \nu b^{\dagger} b +\hbar \frac{\omega_a}{2}\sigma_z - \hbar \lambda (a+a^{\dagger})(b+b^{\dagger})+ H_{e},
\end{equation}
where 
$\lambda =   k e V_{0} C_{0} (mM\nu\omega)^{-1/2} / d^{3}$
is the controllable coupling between the two phononic modes, $\sigma_z = |e\rangle \langle e|-|g \rangle \langle g|$ is the free qubit operator;  
$a,(a^{\dag})$ are the anihilation (creation) operators of the NEM; and $b,(b^{\dag})$ that of the anihilation (creation) operators of mechanical oscillation of the ion; $H_{e}$ describes the interaction of an external laser with the ion
\begin{align*}
H_{e}=-\vec{\mu}\cdot\vec{E}(x,t),
\end{align*}
 
 where $\vec{\mu}$ is the electric dipole operator for the internal
transition.  The electric field describing the laser can be written as
\begin{align}
E(x,t)&=A (e^{i (k_l x-\omega_lt)}+e^{-i (k_l x-\omega_lt)})\nonumber\\
&= A (e^{i [\eta(b+b^{\dagger})-\omega_lt]}+e^{-i [\eta(b+b^{\dagger})-\omega_lt]}) 
\nonumber
\end{align}
where $k_l$ and $\omega_l$ are the wave vector and the frequency of the laser, and $\eta=k_l\sqrt{\frac{\hbar}{2m \nu}}$ is the Lamb-Dicke parameter. The Hamiltonian $H_e$ can be written as:
 
\begin{equation}
H_{e}=\hbar \Omega [\sigma_+ e^{i(\eta(b+b^{\dagger})-\omega_l t)}+\sigma_- e^{-i(\eta(b+b^{\dagger})-\omega_l t)}].
\nonumber
\end{equation}
Where $\Omega=epA/\hbar$ is the effective transition Rabi frequency and $p$ is the dipole matrix element. The qubit dipole transition operators are $\sigma_{-}= |g\rangle \langle e|$ and $\sigma_{+}=\sigma_{-}^{\dag}$.

In the Lamb-Dicke limit, we assuming the ion localized in a region much smaller than the wavelength of the laser $\eta = k_{l}\sqrt{\frac{\hbar}{2m\nu}}\ll 1$, we can express the dipolar interaction between the ion and the field as
 
\begin{equation}\label{eq4}
H_{e}=\hbar\Omega[\sigma_{+}e^{-i\omega_{l}t}+i\eta\sigma_{+}(b+b^{\dagger})e^{-i\omega_{l}t}+H.c.],
\end{equation}

The Hamiltonian  \eqref{eq3} can be written as:
\begin{equation}
H= H_o+H_{I}+H_{e},
\end{equation}
where
\begin{align*}
H_{0}&= \hbar \omega a^\dagger a+ \hbar \nu b^\dagger b +\hbar \frac{\omega_a}{2}\sigma_z,\\
H_{I}&=  - \hbar \lambda (a+a^{\dagger})(b+b^{\dagger}),\\
H_{e}&=  \hbar \Omega [\sigma_{+} e^{-i\omega_l t}+ \sigma_{-} e^{i\omega_l t}]\nonumber\\
&+ \hbar  [ig \sigma_+ (b+b^{\dagger})e^{-i\omega_l t} - ig^{*}\sigma_{-} (b+b^{\dagger})e^{i\omega_l t}].
\end{align*}
with $g=\eta \Omega$. In the interaction picture, with $U(t)=e^{-iH_{0}t/\hbar}$, we have
\begin{align*}\label{eq9}
\mathcal{V}(t)&= U^{\dagger}(t)\big(H_{I}+H_e \big) U(t)\\
&=-\hbar \lambda \big(ab e^{-i(\omega+\nu) t}+ a b^{\dagger} e^{-i(\omega-\nu) t}\\
&+ a^{\dagger}b e^{i(\omega-\nu) t}+  a^{\dagger}b^{\dagger} e^{i(\omega+\nu) t}\big) \\
&+\hbar \Omega [\sigma_{+} e^{-i(\omega_{l}-\omega_{a}) t}+\sigma_{-} e^{i(\omega_{l}-\omega_{a}) t}] \\ 
&+\hbar \big[ig \sigma_{+} b e^{i(\omega_{a}-\omega_{l}-\nu) t} + ig \sigma_{+} b^{\dagger} e^{i(\omega_{a}-\omega_{l}+\nu) t} \\
&- ig^{*} \sigma_{-} b e^{-i(\omega_a-\omega_l+\nu) t} - ig^{*} \sigma_{-} b^{\dagger} e^{-i(\omega_{a}-\omega_{l}-\nu) t}\big].
\end{align*}
Assuming that the frequencies of the NEM and the mechanical oscillation of the ion are nearly in resonance ($\omega+\nu \gg \left| \omega-\nu \right|$), and adjusting the laser frequency for $\omega_{l}=\omega_{a}-\nu$, 
\begin{align*}
\mathcal{V}(t)
&=-\hbar \lambda \big(ab e^{-i(\omega+\nu) t}+ a b^{\dagger} \\
&+ a^{\dagger}b +  a^{\dagger}b^{\dagger} e^{i(\omega+\nu) t}\big) \\
&+\hbar \Omega [\sigma_{+} e^{-i\nu t}+\sigma_{-} e^{i\nu t}] \\ 
&+\hbar \big[ig \sigma_{+} b + ig \sigma_{+} b^{\dagger} e^{2i\nu t} \\
&- ig^{*} \sigma_{-} b e^{-2i\nu t} - ig^{*} \sigma_{-} b^{\dagger} \big].
\end{align*}


In the RWA we have
\begin{equation}
\mathcal{H} \approx \hbar\frac{\omega}{2} \sigma_{z} + \hbar\omega a^{\dag}a + \hbar\omega b^{\dag}b - \hbar \lambda \left(ab^{\dagger}+ba^{\dagger} \right) + \hbar g \left( \sigma_{+}b+\sigma_{-}b^{\dagger} \right).
\end{equation}
where $\omega=\nu$, $ib\to b$ ($-ib^{\dag}\to b^{\dag}$) with Ramsey rotation.

Adding a drive with amplitude $|\epsilon|$ and frequency $\omega_{p}$ to NEM
\begin{eqnarray*}
    H_{d} = \hbar ( \epsilon a^{\dag}e^{-i\omega_{p}t} + \epsilon^{*} a e^{i\omega_{p}t} ),
\end{eqnarray*}
the full Hamiltonian results
\begin{eqnarray}
\mathcal{H}_{Full} &=& \hbar\frac{\Delta_{p}}{2} \sigma_{z} + \hbar\Delta_{p} a^{\dag}a + \hbar\Delta_{p} b^{\dag}b - \hbar \lambda \left(ab^{\dagger}+ba^{\dagger} \right) \nonumber\\
&+&\hbar g \left( \sigma_{+}b + \sigma_{-}b^{\dagger} \right)
+ \hbar ( \epsilon a^{\dag} + \epsilon^{*} a  ),
\end{eqnarray}
where $\Delta_{p}=\omega-\omega_{d}$.

The presence of dissipation in the circuit dynamics can be described by the master equation \cite{carmichael2009open}
\begin{equation}
\partial_{t} \hat{\rho} = {-i\hbar} \left [ H_{Full},  \hat{\rho} \right ]
+ \frac{\gamma}{2} \mathcal{D}[ \sigma_{-}]	\hat{\rho}
+ \frac{\gamma_{\phi}}{2} \mathcal{D}[ \sigma_{z}]	\hat{\rho}
+ \sum_{\alpha=a,b}
\frac{ \kappa_{\alpha}}{2} \mathcal{D}[\alpha] \hat{\rho}
\end{equation}
where 
$\mathcal{D}[\beta] \hat{\rho} = 2 \beta \hat{\rho} \beta^{\dag}-\left\{ \beta^{\dag}\beta,\hat{\rho} \right\}$, for $\beta=\sigma_{-},\sigma_{z},\alpha$,
with $\kappa_{a}$, $\kappa_{b}$ and $\gamma$ being the NEM and QuBit relaxation rates, respectively. The $\gamma_{\phi}$ is the QuBit decoherence rate. With this we can obtain the following equations of motion
\begin{subequations}\label{eqm}
\begin{equation}
	\left\langle \dot{a} \right\rangle = -i\bar{\Delta}_{a} \left\langle a \right\rangle -i\epsilon + i \lambda \left\langle b \right\rangle  ,
\end{equation}
\begin{equation}
\left\langle \dot{b} \right\rangle = -i\bar{\Delta}_{b} \left\langle b \right\rangle + i \lambda \left\langle a \right\rangle -  ig \left\langle \sigma_{-} \right\rangle ,
\end{equation}
\begin{equation}
	\left\langle \dot{\sigma}_{-} \right\rangle = -i\bar{\Delta}_{q} \left\langle \sigma_{-} \right\rangle + ig \left\langle b \sigma_{z} \right\rangle  ,
\end{equation}
\end{subequations}
where $\kappa_{q}/2=\gamma_{\phi}+\gamma/2$ and $\bar{\Delta}_{j}=\Delta_{p}-i\kappa_{j}/2$, for $j=a,b,q$. 
In single phonon limit, we have $\left\langle b\sigma_{z} \right\rangle=-\left\langle b \right\rangle$. With this we have the following solutions in the steady state
\begin{subequations}\label{ss}
\begin{equation}
	\left\langle a \right\rangle_{ss} = \frac{\epsilon \left( \bar{\Delta}_{b}\bar{\Delta}_{q} - g^{2} \right)  }{g^{2} \bar{\Delta}_{a} + \lambda^{2}\bar{\Delta}_{q} - \bar{\Delta}_{a}\bar{\Delta}_{b}\bar{\Delta}_{q}} ,
\end{equation}
\begin{equation}
\left\langle b \right\rangle_{ss} = \frac{\epsilon \lambda \bar{\Delta}_{q}}{g^{2} \bar{\Delta}_{a} + \lambda^{2}\bar{\Delta}_{q} - \bar{\Delta}_{a}\bar{\Delta}_{b}\bar{\Delta}_{q}} ,
\end{equation}
\begin{equation}
	\left\langle \sigma_{-} \right\rangle_{ss} =  \frac{\epsilon \lambda g }{\bar{\Delta}_{a}\bar{\Delta}_{b}\bar{\Delta}_{q} - g^{2} \bar{\Delta}_{a} - \lambda^{2}\bar{\Delta}_{q} } ,
\end{equation}
\end{subequations}
For we investigate the phenomenon of induced transparency in this paper, we'll call it  nanomechanically induced transparency (NIT). It is important to highlight that the occurrence of the NIT phenomenon requires that $\kappa_{a} \gg \kappa_{b}, \gamma, \gamma_{\phi}$. 
To plot the graphs we can use experimental data recorded in the literature, given with $\lambda\approx 300 kHz$ \cite{hensinger05}, and $g\approx 500 kHz$ \cite{rmp2003} for $Cd^{+}$, and $\kappa_{q}=11.9 kHz$ \cite{meekhof1996generation}. The other parameters can be modified by the manufacturing technology of the experimental elements. Here we will only show their implications. Next, we will investigate the effect of transmission and absorption due to the oscillatory mechanical parameters analogous to the electromagnetic parameters of Ref. \cite{souza2015electromagnetically,diniz2018multiple}.

\begin{center}
\begin{figure}[h]
\begin{center}
\includegraphics[scale=0.4]{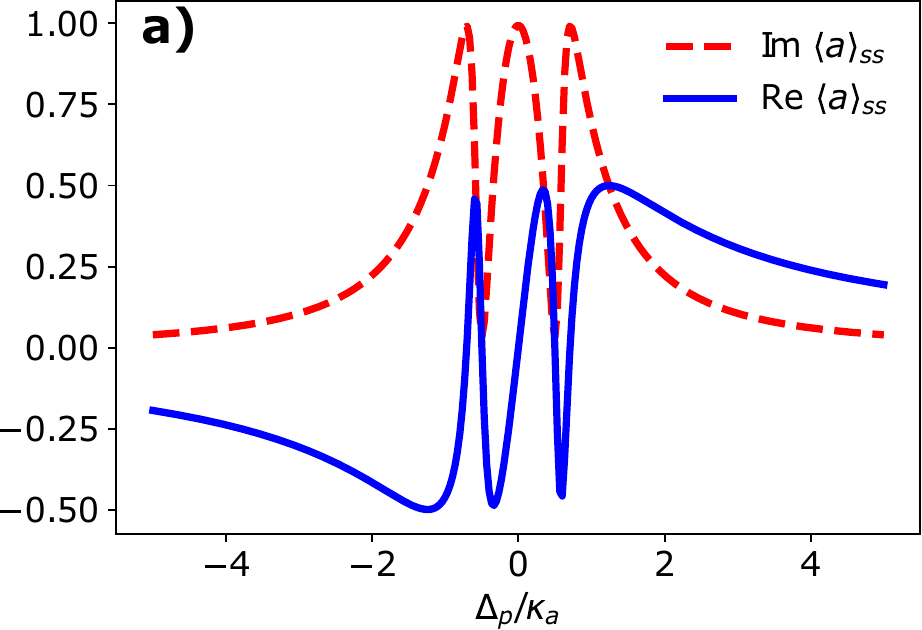}
\includegraphics[scale=0.4]{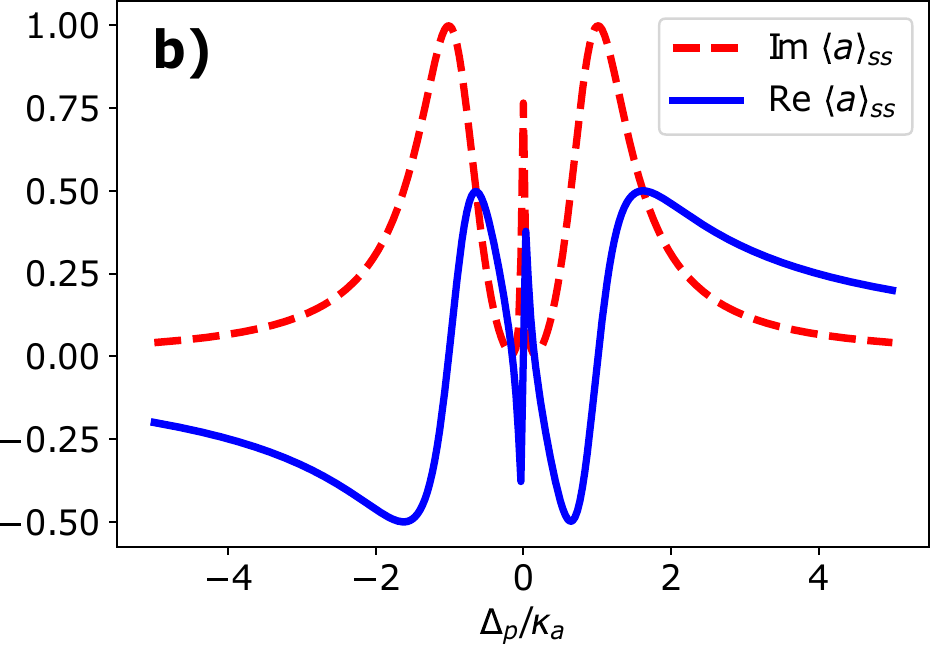}
\caption{Absorption $ Im \left \langle a \right \rangle_{ss}$ (red dashed line) and dispersion $ Re \left \langle a \right \rangle_{ss}$ (blue solid line) of the cavity mode $a$ when coupled to trapped ion and cavity mode $b$ as a function of the normalized detuning $\Delta_p/\kappa_a$. In this case we use the parameters: a) $\left |\epsilon  \right | = 0.03\kappa_a, \lambda = g = 0.5 \kappa_a, \kappa_b = 10^{-3}\kappa_a,  \gamma = \gamma_\phi =10^{-3} \kappa_a$; b) $\left |\epsilon  \right | = 0.03\kappa_a,  \lambda = 1.0\kappa_a,  g = 0.15 \kappa_a,  \gamma = \gamma_\phi =10^{-3} \kappa_a $. }
\label{fig2}
\end{center}
\end{figure}
\end{center}

\begin{center}
\begin{figure}[h]
\begin{center}
\includegraphics[scale=0.35]{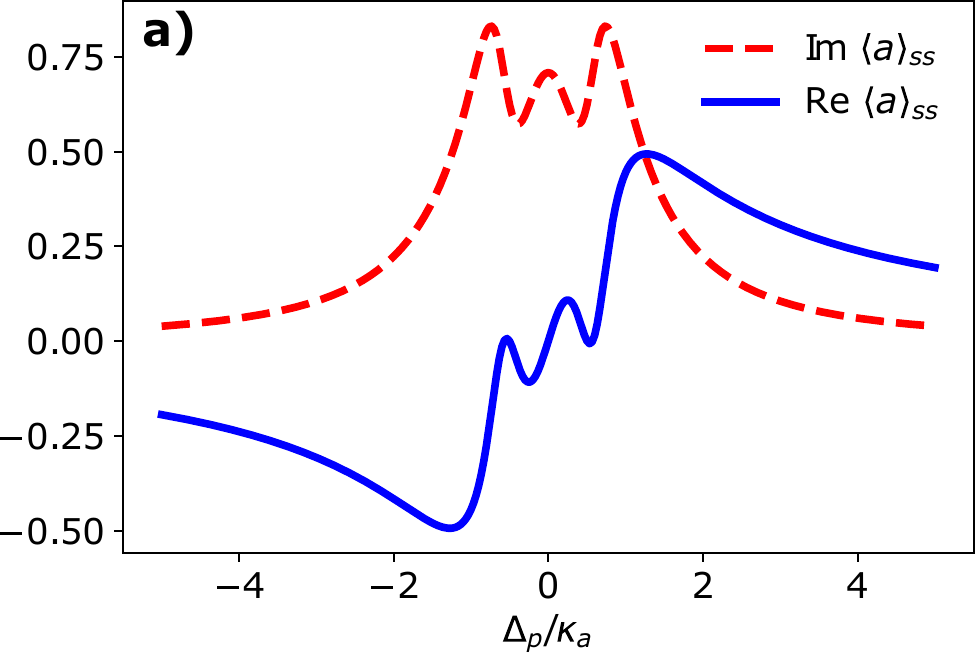}
\includegraphics[scale=0.35]{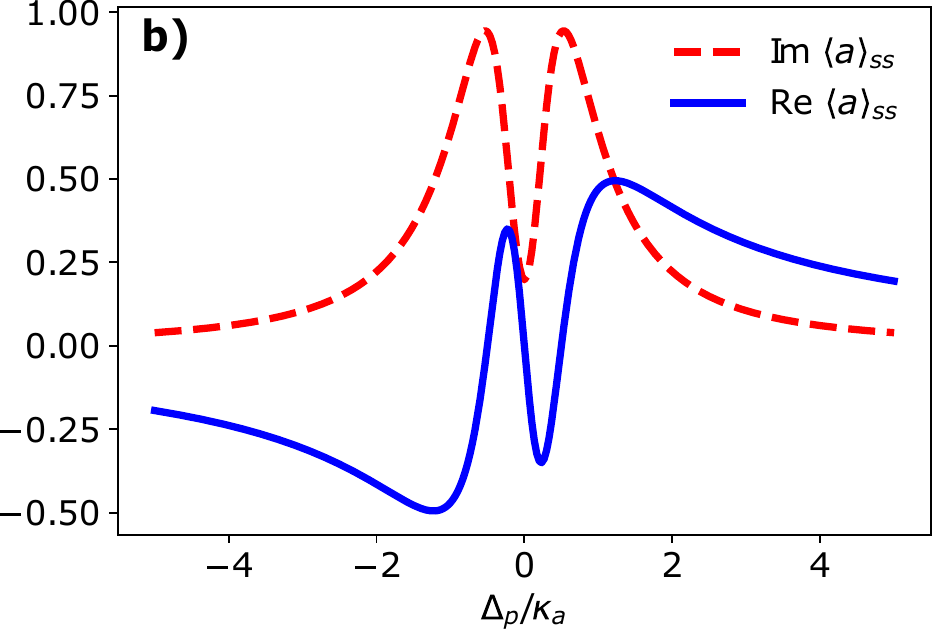}
\caption{Absorption $ Im \left \langle a \right \rangle_{ss}$ (red dashed line) and dispersion $ Re \left \langle a \right \rangle_{ss}$ (blue solid line) of the cavity mode $a$ when coupled
to trapped ion and cavity mode $b$ as a function of the normalized detuning $\Delta_p/\kappa_a$. In this case we use the parameters: a) $\left |\epsilon  \right | = 0.03\kappa_a, \lambda = g = 0.5 \kappa_a, \kappa_b = 10^{-3}\kappa_a,  \gamma = 10^{-2} \kappa_a, \gamma_\phi =10^{-1} \kappa_a$; b) $\left |\epsilon  \right | = 0.03\kappa_a, \lambda = g = 0.5 \kappa_a, \gamma = 10^{-1} \kappa_a, \gamma_\phi = 1.0\kappa_a $. }
\label{Fig3}
\end{center}
\end{figure}
\end{center}

\section{Results and Discussion}
In Fig. (\ref{fig2}), we show the behavior of the real and imaginary parts of the operator $\left \langle a \right \rangle_{ss}$ given by the Eq.(\ref{ss}a). 
As can be seen in Fig (\ref{fig2}.a), when the couplings are equal, the transparency windows have a symmetrical profile. However, when there is a discrepancy between the coupling values, the transparency windows can be significantly affected, as can be seen in Fig. (\ref{fig2}.b). This fact is discussed in Ref. \cite{diniz2018multiple} and is related to the interference between the different absorption paths involving the closest energy levels that are in the linewidth of the system. 
We emphasize that the width and height of the central peak in Fig. (\ref{fig2}) are strongly dependent on the transition rates between the ground state and the first excited states. Specifically, these rates can be equal at certain values of g. In such cases, selecting values of g where these rates are equal can result in different absorption profiles for the system. For instance, by choosing $\left ( g \sim 0.15 \right )$, we obtain the absorption profile depicted in Fig. (\ref{fig2}.b). Indeed, increasing the value of g leads to an increase in both the height and width of the central peak, as observed in Fig. (\ref{fig2}.a).
In this way, we have a system that can be controlled by adjusting the system coupling parameters.
Furthermore, in this system it is possible to have Fano-type resonance by adjusting the coupling parameters. This type of resonance arise of the interference of eigenstates whose energy levels are partially inside and outside the linewidth of the system. For example, for the case ($g = 3.0\lambda$), the absorption exhibits an asymmetric shape.
In comparison to the study described in Ref.\cite{yuan2008nanomechanical}, we show that it is possible to obtain a broader scenario of the system’s absorption profile turning on the coupling parameters of the system.
We also observed that dissipative ion channels can substantially modify the absorption profile of the system as shown in Fig. (\ref{Fig3}.a). As can be seen in Fig. (\ref{Fig3}.b), the dephasing dissipative channel is what most damages the system's absorption. In fact, it destroys the central absorption peak, providing a strong indication of a destructive interference effect on the system.  

\section{Summary} 
We show how the effects of ion degrees of freedom in the Lamb-Dicke approximation can be relevant in the emergence of quantum interference phenomena in the context of nanoelectromechanical systems. The nanomechanically induced transparency (NIT) phenomenon arises due to interference of the absorption paths that are in the linewidth of the system. We also show that the profile of the absorption spectrum can be controlled by tuning the coupling parameters between the ion and the nanoelectromechanical resonators modes. The results presented can be pave the way for the intense investigations of the analog electromagnetically induced transparency in nanoscale. 

{\it Acknowledgments.} 
E. C. Diniz acknowledge the UNEMAT (Campus Tangará da Serra) for the hospitality.

\nocite{*}
\bibliography{aipsamp}


\end{document}